\documentclass[letterpaper,twocolumn]{jpsj3}
\usepackage{txfonts}

\usepackage{siunitx}
\usepackage{textcomp}

\usepackage{algpseudocode}
\usepackage{algorithm}

\usepackage{here}

\usepackage{tabularx}

\usepackage{graphicx}
\usepackage{float}
\usepackage{textcomp}
\usepackage{amsmath}
\usepackage{listings}
\usepackage{url}

\usepackage{tikz}
\usetikzlibrary{trees}

\title{Efficient Algorithm for Binary Quadratic Problem \\by Column Generation and Quantum Annealing}

\author{Sota Hirama and Masayuki Ohzeki}
\inst{Graduate School of Information Sciences, Tohoku University, Miyagi 980-8564, Japan \\
Department of Physics, Tokyo Institute of Technology, Tokyo, 152-8551, Japan \\
International Research Frontier Initiative, Tokyo Institute of Technology, Tokyo, 108-0023, Japan \\
Sigma-i Co., Ltd., Tokyo, 108-0075, Japan
} 

\abst{
We propose an efficient algorithm that combines column generation and quantum annealing to solve binary quadratic problems. 
Binary quadratic problems are difficult to solve because they are NP-hard.
An attempt to solve binary quadratic problems efficiently by column generation has been studied, but it demands successively solving quadratic unconstrained binary optimization problems. 
We solve the bottleneck by using quantum annealing or simulated annealing. 
Our results demonstrate a good approximate solution obtained in 2.7 to 1000 times
shorter computational time to use column generation and quantum annealing to solve binary quadratic problems than the existing fast solver.
}

\begin{document}
\maketitle
\renewcommand{\algorithmicrequire}{\textbf{Input:}}
\renewcommand{\algorithmicensure}{\textbf{Output:}}

{\it Introduction.}
Quantum annealing (QA) is known to be a method for solving generic combinatorial optimization problems \cite{Kadowaki1998, Ohzeki2015}.
In particular, its physical realization, a quantum annealer, is expected to be a quick solver for the quadratic unconstrained binary optimization (QUBO) problems, which can be implemented therein.
Various applications of QA are proposed as in traffic flow optimization\cite{neukart2017traffic,hussain2020optimal,inoue2021traffic}, 
 finance \cite{rosenberg2016solving, orus2019forecasting, venturelli2019reverse}, logistics \cite{feld2019hybrid,ding2021implementation}, manufacturing \cite{venturelli2016quantum, Yonaga2022, Haba2022}, preprocessing in material experiments\cite{Tanaka2023}, marketing \cite{nishimura2019item}, steel manifacturing \cite{Yonaga2022}, and decoding problems \cite{IdeMaximumLikelihoodChannel2020, Arai2021code}.
The model-based Bayesian optimization is also proposed in the literature \cite{Koshikawa2021}
A comparative study of quantum annealer was performed for benchmark tests to solve optimization problems \cite{Oshiyama2022}. 
The quantum effect on the case with multiple optimal solutions has also been discussed \cite{Yamamoto2020, Maruyama2021}. 
As the environmental effect cannot be avoided, the quantum annealer is sometimes regarded as a simulator for quantum many-body dynamics \cite{Bando2020, Bando2021, King2022}. 
Furthermore, applications of quantum annealing as an optimization algorithm in machine learning have also been reported \cite{neven2012qboost,khoshaman2018quantum,o2018nonnegative, Amin2018,Kumar2018,Arai2021,Sato2021,Urushibata2022,hasegawa2023}.

Unfortunately, due to its limitation, the current quantum annealer can not efficiently solve the combinatorial optimization problem with equality/inequality constraints, even by using the various techniques, at a satisfactory level \cite{Ohzeki2020, Yonaga2020}.
Most of the combinatorial optimization problems in practice implement various types of constraints.
In the classical computer, the combinatorial optimization problem is efficiently solved by various algorithms.
Elaborate classical algorithms can also inspire quantum computation.
In the present study, we propose the combination of QA with a classical method for solving combinatorial optimization problems with a large number of constraints, namely column generation \cite{Barnhart1998}.
Our results demonstrate that our method can solve efficiently constrained combinatorial optimization problems.

{\it Problem setting.}
We solve the following type of constrained binary optimization problem:
\begin{equation}
\min_{\bf x}\left\{ \sum_{ij}Q_{ij} x_i x_j \right\}~{\rm s.t.}~ \sum_{ij} A_{kij} x_i x_j \leq b_k~\forall k ,  x_i\in\{0, 1\}~\forall i,
\end{equation}
where $Q_{ij}$ is the continuous-valued element of the QUBO matrix designing the cost function to be solved, $A_{kij}$ is the continuous-valued element of the matrix, and $b_i$ is the continuous-valued element of the vector, which define the equality/inequality constraints.
The variable $x_i$ is binary as $0$ or $1$. 
The number of variables is denoted as $n$.
We find a solution that minimizes the cost function while adhering to the constraints.
When we input the constrained optimization problem into the quantum annealer, we generally modify the cost function to describe the constraints as in the penalty method.

{\it Column generation.}
In our study, we instead utilize column generation, a popular algorithm for the constrained optimization problem \cite{Barnhart1998}.
In particular, it efficiently solves the large-size linear programming problem with continuous variables.
We thus convert a quadratic problem into a linear programming problem to adapt column generation. 
We use several techniques in convex optimization problems.
We define a convex combination as a linear combination of extreme points by considering a convex hull for some set.
\begin{align}
    x_a = \sum_{p \in \mathcal{P}} x^{p}\lambda^p, {\rm s.t.} \sum_{p \in \mathcal{P}} \lambda^p = 1, \lambda^p \geq 0, \; \forall p
\end{align}
where $x^k$ is an extreme point and $\mathcal{P}$ is the index set of all the possible extreme points. 
We use a convex combination to transform the binary quadratic programming problem. 
We iteratively solve the effective optimization problem later and then obtain several solution vectors.
We take a convex hull of the solution-vector set. 
First, we define a quadratic problem converted by convex combination\cite{Bettiol2019}. 
\begin{align}
    &\min_{\bf \lambda} \sum_{p\in\mathcal{P}} \sum_{ij}Q_{ij} x^p_i x^p_j \lambda^p, \\
    &{\rm s.t.}\;\; \sum_{p\in\mathcal{P}} \sum_{ij} A_{k ij} x^p_i x^p_j \lambda^p \leq b_i,\quad~\forall k \\
    &\quad\;\; \sum_{p\in \mathcal{P}} \lambda^p = 1,\\
    &\quad\;\; \lambda^p \geq 0,\quad~\forall p \in \mathcal{P}
\end{align}
where $x^p\in \{0, 1\}^n$ are binary constant vectors and $\mathcal{P}$ is the index set of all the possible extreme points of the solution-vector set.
Here we set the variables $\lambda^p$.
We utilize the column generation method on this quadratic problem.

The column generation method is an algorithm that efficiently finds the optimal solution by starting the search from the minimum necessary extreme points and generating extreme points until the optimal solution is reached.
The computational time is highly reduced than considering the extreme points of the convex hull of all solution sets. 

We use the Dantzig-Wolfe Decomposition to modify the quadratic problem by changing $\mathcal{P}$ into its subset $\bar{\mathcal{P}}$, the restricted set of extreme points. 
We call this the restricted master problem (RMP). 

We here consider a dual problem of RMP. 
The dual problem of linear programming problems can be obtained by interchanging the coefficients of the objective function and the right-hand side of the constraint.
\begin{align}
    &\max_{\bf \rho, \bf \pi_0} \sum_{i} b_i\rho_i + \pi_0,\\
    &{\rm s.t.}\;\; \sum_{k=1}^{m} \sum_{ij} A_{k ij} x^p_i x^p_j \rho_k + \pi_0 \leq \sum_{ij}Q_{ij} x^p_i x^p_j,\quad \forall p\in\bar{\mathcal{P}}\\
    & \quad\;\; \rho \leq 0,
\end{align}
where $\rho \in \mathbb{R}^n$ and $\pi_0 \in \mathbb{R}$ are the dual variables.
there is one dual variable for each explicit constraint in RMP. 
A solution to a dual problem provides a good lower bound on RMP.
If there are points $x_p\in\mathcal{P}\setminus\bar{\mathcal{P}}$, the answer of RMP is not optimal. 

{\it Central QUBO problem.}
Therefore, the main issue is finding a solution that lowers the cost function of RMP.
We define a pricing problem that adds the new extreme point to $\bar{\mathcal{P}}$ lowering the cost function of RMP. 
By solving the dual problem, we obtain the dual variables $\rho^*$ and $\pi_0^*$. 
We use these dual variables and define the pricing problem by
\begin{align}
    &\min_{\bf x} \sum_{ij}\left\{\left( Q_{ij}-\sum_{k=1}^{m}\rho^*_k A_{kij} \right) x_i x_j\right\} - \pi_0^*,\\
    &{\rm s.t.}\;\; x\in\{0,1\}^n,
\end{align}
where $\rho^* $ and $\pi_0^*$ are the solution of dual problem.
The objective function expresses the residual characterizing discrepancy in the constraints.
One can find that this problem takes the form of the quadratic unconstrained binary optimization (QUBO).\\

It is noticed that the resulting solution is approximate, depending on the precision to solve the QUBO problem.
However, the main issue is reduced to increasing the lower bound of the original optimization problem (or decreasing the upper bound in the case of a maximization problem).
One can assess the quality of the tentative solution while proceeding with the computation.

The point is to transform the original constrained optimization problem into the unconstrained one.
The quantum annealer can efficiently solve the QUBO problem even at the current level.
We incorporated simulated annealing (SA) \cite{Kirkpatrick1983} and QA into solving the pricing problem.
Below, we investigate the performance of our method.

{\it Results.}
We solved randomly generated problems using our proposed method and compared its performance to Gurobi, one of the fastest commercial solvers, as a benchmark.
The version of Gurobi used in this experiment was 10.0.1.

We generated our test problems under the following conditions : $Q_{ij}\in\{-1, 1\}(i \leq j)$, $A_{kij}\in\{-1, 1\}(i\leq j)$ and $b_k = 1$. 
We prepare 100 instances for each problem size.

First, we use QA to solve the pricing problem. 
We use D-Wave quantum annealer, 2000Q, and Advantage.
The annealing time for QA was set to 500\textmu s.
We include the communication time from Japan to Canada as the real computation time for QA.
Two types of solutions using Gurobi are used for comparison.
The exact solution is denoted by Gurobi, and the approximate solution is denoted by R-Gurobi up to the value of the result derived by QA.
We set the calculation time limit as 1800 seconds (30 minutes) for R-Gurobi.

\begin{figure}[t]
\begin{center}
\includegraphics[width=80mm]{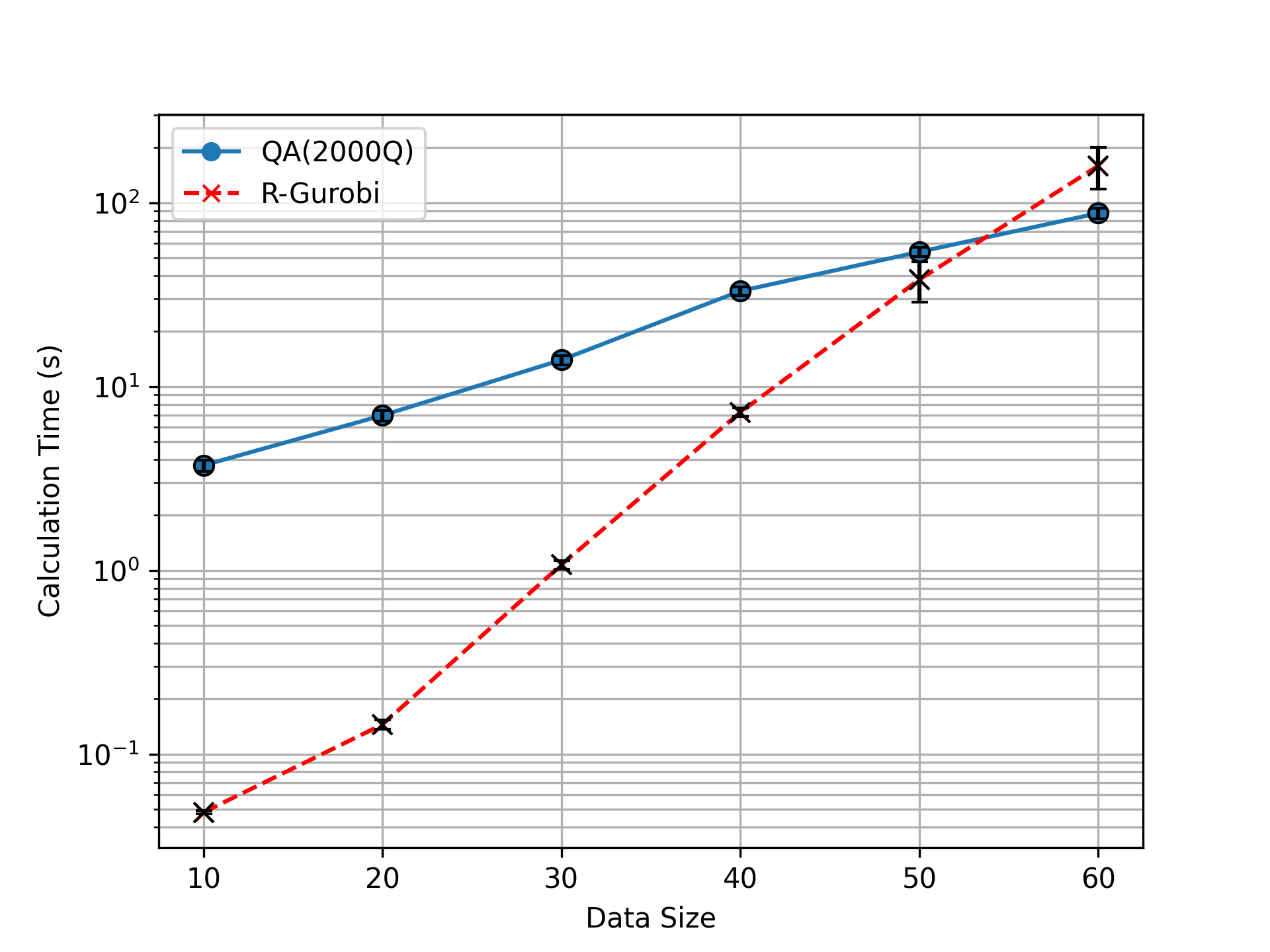}
\includegraphics[width=80mm]{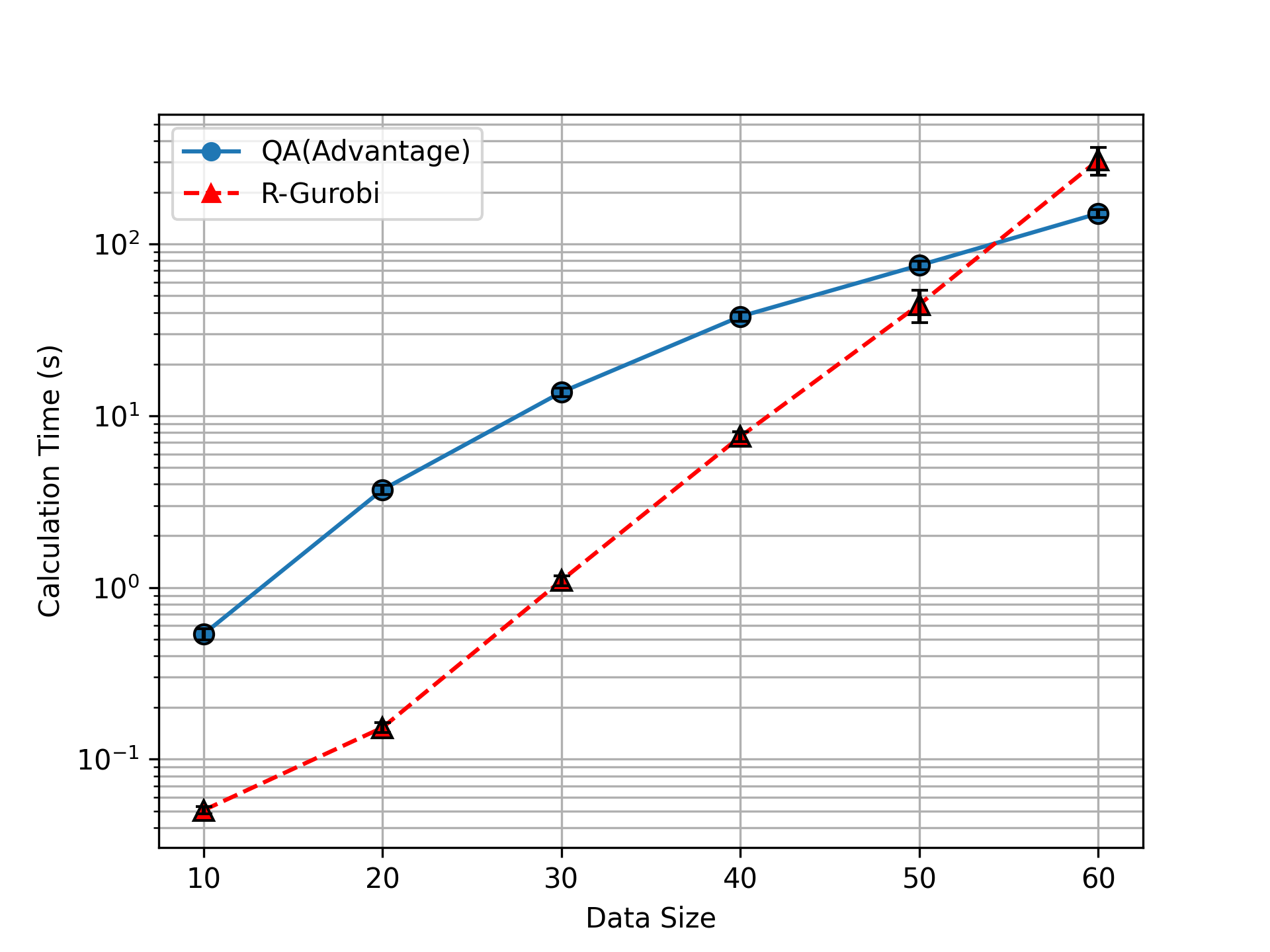}
\caption{Relationship between problem size and computation time up to the same quality of the solution in QA(Advantage) and R-Gurobi}
\label{res3}
\end{center}
\end{figure}

Next, we discuss the result between our algorithm and R-gurobi. The QA (2000Q) and R-Gurobi result is shown in Fig. \ref{res3}.
The comparison results are shown in Fig. \ref{res3} shows R-Gurobi is faster than QA (2000Q) and QA (Advantage) up to 50. 
However, when the problem size reaches 60, both QAs reach an approximate solution faster.
It is often said that the quantum annealer can efficiently solve small-size problems.
However, our method overcomes the Gurobi using the advantage to solve the binary quadratic problem, which finds the extreme points.
The bottleneck of QA is how to deal with the constraints.
One usually uses the penalty method, which demands a large coefficient value in the cost function, resulting in the decaying precision of the solution.
We avoid the penalty method and instead use the column generation while using the QUBO.

\begin{figure}[t]
\begin{center}
\includegraphics[width=90mm]{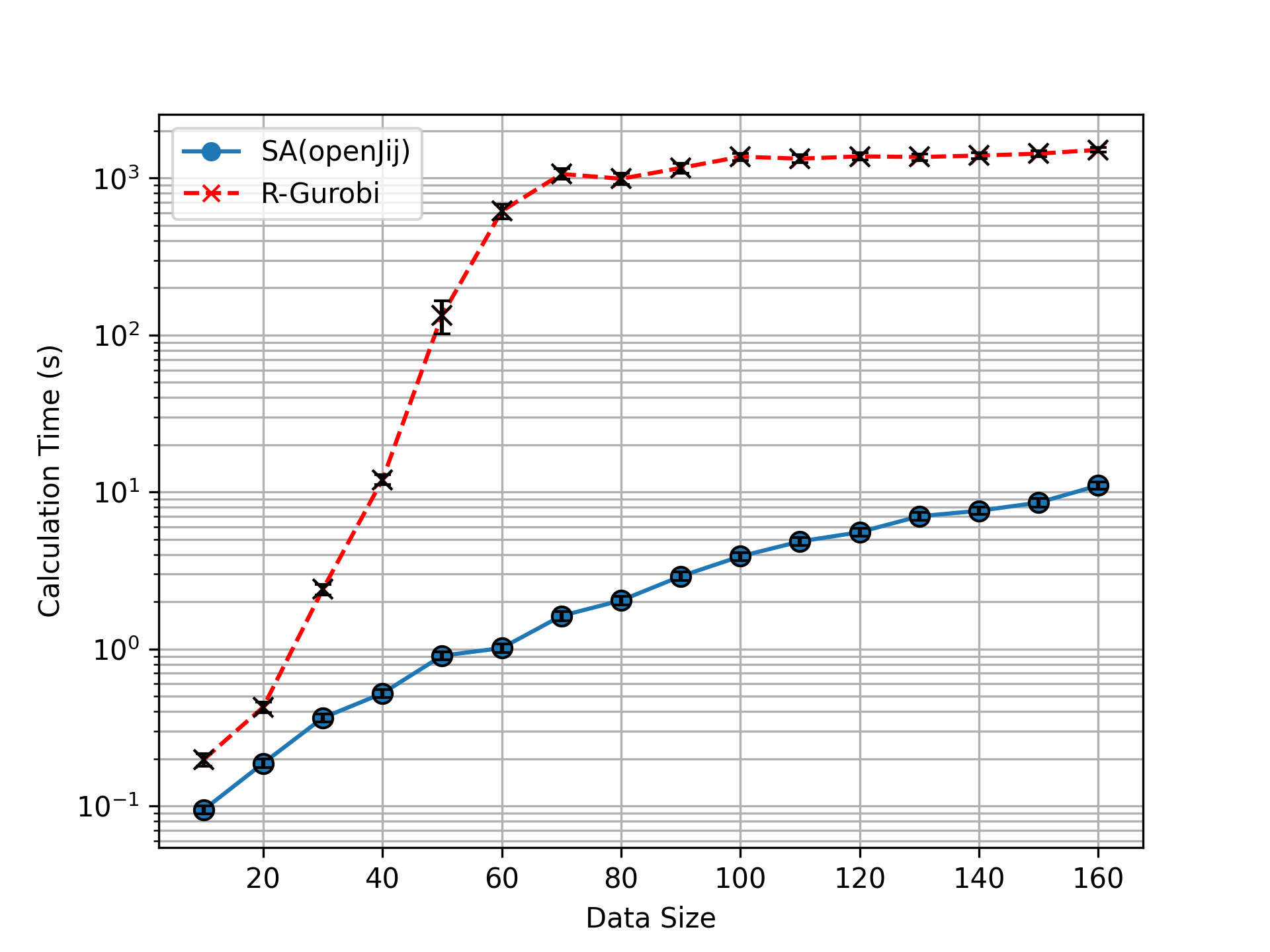}
\caption{Relationship between problem size and calculation time (problem size up to 160).
The computation time of R-Gurobi takes highly longer but our method shows a moderate increase of the computation time.
The largest size 160 is the limitation of the computation by using the Gurobi in our computation environment. }
\end{center}
\label{res5}
\end{figure}

Unfortunately, the precision of the solution by QA at the current level is worse than SA on the classical computer in general.
We test our method by SA hereafter.
We use OpenJij to perform SA. 
The parameters of SA (OpenJij) were used as default settings.
We can test our method for larger-size problems up to 160 in the classical computer.
The result of SA (OpenJij) and Gurobi is shown in Fig. \ref{res5}.

The results show that for all the problem sizes, our method using SA completes the computation faster than R-Gurobi.
We compute the ratio time of QA (2000Q, Advantage) and SA to R-Gurobi as in Table \ref{table1}.

\begin{table}[t]
\caption{Relationship between problem size and rate of calculation time. 
The parenthesis denotes that the result is just an estimation because we use the results when R-Gurobi ceases its computation due to time limitations.
}
\label{table:data_type}
\centering
\begin{tabular}{cccc}
\hline
problem size & QA(Advantage)  &  QA(2000Q) & SA  \\
\hline
10 & 0.1297 & 0.01970 & 2.783 \\

20 & 0.05732 & 0.02909 & 3.234 \\

30 & 0.1168 & 0.01970 & 10.50 \\

40 & 0.2797 & 0.3027 & 30.51\\

50 & 0.7328& 0.8942 & 220.3\\

60 & 3.750 & 2.914 & 836.0\\

70 & - & - & (1001)\\
\hline
\end{tabular}

\label{table1}
\end{table}
The results show how faster QA and SA can find solutions than R-Gurobi.
The results (Table \ref{table1}) show that QA (2000Q) is 3.75 times faster in finding a solution when the problem size is 60, while Advantage is 2.914 times faster when the problem size is 60. 
SA takes the maximum ratio value when the problem size is 70 and can find the solution 1001 times faster than R-Gurobi.
However, the R-Gurobi tends to cease its computation due to a time limitation of up to 30 minutes.
Thus the potential of our method to speed up the large problem size would be more impact.

\begin{figure}[t]
\begin{center}
\includegraphics[width=80mm]{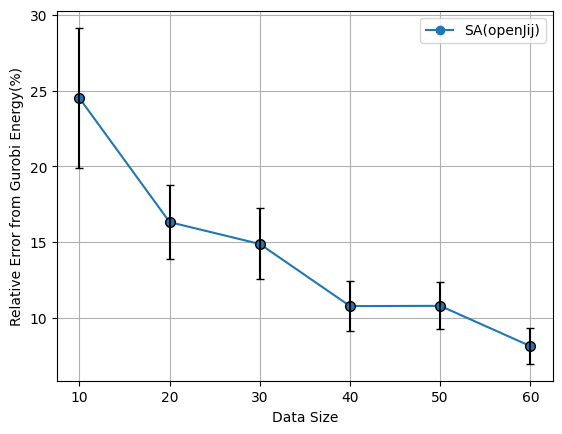}
\includegraphics[width=80mm]{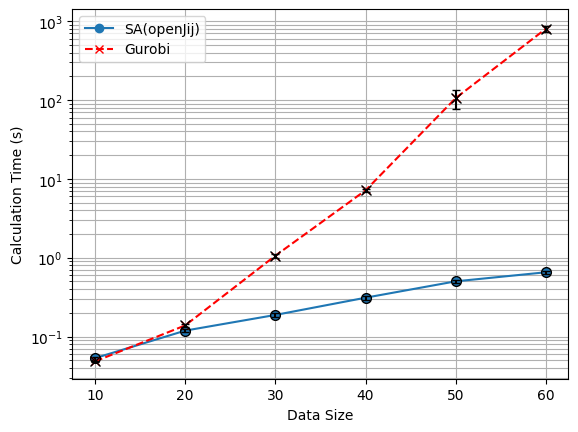}
\caption{Relationship between problem size and residuals from exact solutions, and computation time in SA and Gurobi}
\label{res1}
\end{center}
\end{figure}

We investigate the precision of our method in comparison with the exact solver, Gurobi, as in Fig. \ref{res1}.
The results show that the larger the problem size, the smaller the approximate solution's energy deviates from the exact solution's energy.
This is because the large size means an increase of the constraints in our tests.
Thus column generation works well in large-size cases.

{\it Conclusion}
In this letter, we have verified that the column generation algorithms introducing QA and SA efficiently solve binary quadratic programming problems with several constraints. 
The column generation with QA(2000Q) was up to 3.750 times faster than R-Gurobi, QA(Advantage) was up to 2.914 times faster than R-Gurobi, and SA was potentially up to 1001 times faster than R-Gurobi. 
We consider why this algorithm can find an approximate solution faster than Gurobi: Gurobi is an algorithm that tries to find an exact solution by iterative calculations using branch-and-bound and other methods, while our algorithm tries to derive an approximate solution by using a column generation method. 
The time-consuming part of our algorithm is the pricing problem. 
Since we could reduce the computation time using QA or SA, we could derive an approximate solution faster than Gurobi.
The QUBO problem, which is hard to solve, demands an exponentially long time, depending on the problem size.
Thus most of the algorithms rely on various heuristics and approximations via semi-definite programming.
Our method, in other words, by use of SA and QA is also in the line of this direction to solve the QUBO problem.
Moreover, the reason why SA was able to solve the problem faster than QA is that QA includes the communication time from our point to Canada in the computation time. 
In contrast, SA only needs simple computation time without communicating on the local classical computer.
We here emphasize one of the advantages points of our method.
Our method can solve the quadratic binary optimization problem without the penalty method.
Thus we do not need any parameter tuning to solve the optimization problem with constraints in QA.

It is important to test our method in QPLIB, a library of quadratic programming instances \cite{Furini2019}.
Since our method is based on column generation, it would perform better against the optimization problem with more constraints.
More investigations in this direction would be more important.

{\it Acknowledgement.}
This work is supported by JSPS KAKENHI Grant No. 23H01432.

\bibliographystyle{jpsj}
\bibliography{main}
\end{document}